\documentclass[preprint,showpacs,preprintnumbers,amsmath,amssymb]{revtex4}
\usepackage{graphicx}
\begin{document}

\title{Plasmons in a Superlattice  of Fullerenes or Metallic Shells  }

\author{ Antonios Balassis,$^{\dag}$ and Godfrey Gumbs$^{\ddag,\ast}$ }
\affiliation{$^{\dag}$Department of Physics and Engineering Physics, Fordham University, 441 East Fordham Road,
       Bronx, NY 10458, USA\\
$^{\ddag}$Department of Physics and Astronomy\\
Hunter College of the City University of New York\\
695 Park Avenue, New York, NY 10065\\
$^{\ast}$ Donostia International Physics Center (DIPC),
P. Manuel de Lardizabal 4, 20018 San Sebastian, Spain
}

\begin{abstract}

A theory for the collective plasma excitations in a linear periodic array of
spherical two-dimensional electron gases (S2DEGs)    is presented. This is
a simple model for an ultra thin and narrow microribbon of fullerenes or
metallic shells. Coulomb coupling
between electrons located on the same sphere and on different  spheres is included in
the random-phase approximation (RPA). Electron hopping between spheres is neglected in
these calculations.   The resulting plasmon-dispersion equation is solved numerically.
 Results are presented for a superlattice of single-wall S2DEGs   as a function of the
 wave vector. The plasmon dispersions are obtained for different spherical      separations.
  We show that the one-dimensional translational symmetry of the lattice is maintained in the
plasmon spectrum. Additionally, we compare the plasmon dispersion when the superlatice
direction is parallel or perpendicular to the axis of quantization. However,
because of anisotropy in the Coulomb matrix elements, there is anticrossing in the
plasmon dispersion only in the direction perpendicular to the quantization axis.
The S2DEG may serve as a simple model for fullerenes, when their energy bands are far apart.

\end{abstract}


\maketitle

\section{  Introduction}
\label{sec1}

Research on the properties of carbon-based materials has soared
over the years mainly because it may be found in a variety of
allotropic  forms. These include graphite, graphene,  carbon
nanotubes and nanoscrolls as well as fullerenes. In particular,
the discovery of fullerenes\,\cite{O1,O1b,O2,O3,O4,O5,O6}
has led to several impressive
advances in the study of  carbon  nanoparticles \cite{XXX1}.
The optical properties of fullerenes   have been
investigated from both an experimental and theoretical point of
view \cite{EELS,Apell,expt1,expt2}. Since optical
measurements are a non-invasive probe of the samples, they
provide reliable information about the electronic properties
of fullerenes.

\medskip

\begin{figure}[t]
  \centering
  \includegraphics[width=5.67in,height=3.84in,keepaspectratio]{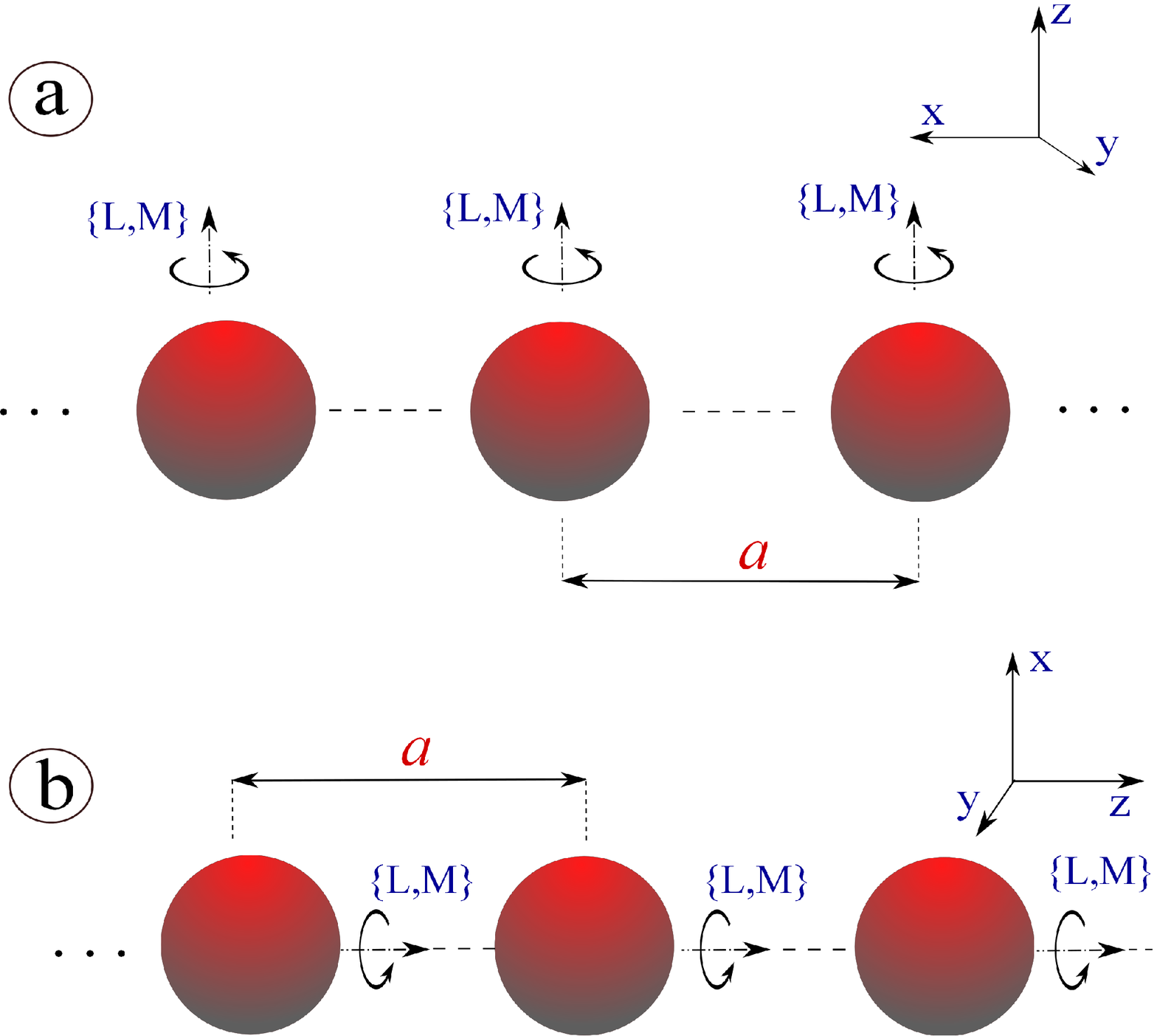}
  \caption{  (Color online) Schematic representation of a superlattice
	of spherical two-dimensional electron gases  (a) perpendicular and (b) along the
	axis of quantization.}
  \label{fig11}
\end{figure}

As a result of  the recent advances in techniques such as solvent-assisted
self-assembly \cite{XXX2}, fullerenes can now be produced in abundant quantities
even to form thin films of fullerene-like MoS     nanoparticles \cite{XXX3}
and have stimulated  renewed interest in these materials.
Additionally, the ability to control optical fields has now made
it feasible to  ascertain the plasmon excitations in pre-arranged
arrays of fullerenes \cite{XXX2,XXX3}.
We consider a simple model for an ultra narrow and thin film of fullerenes
forming a microribbon by employing a one-dimensional superlattice.
   In this work, we  investigate the photoexcitation
of plasmons in a regular periodic array of fullerenes shown schematically
in Fig.\ (\ref{fig11}). Our goal
is to  thoroughly analyze the dependence of the plasma frequency
on the relative orientation of the EM probe field with respect to the
axis along which the periodicity occurs.

\medskip

We demonstrate that the optical absorption spectra of such nano spheres exhibit
a rich dependence on the magnitude and direction of the  transfer momentum.
The wave functions have spatial symmetry originating from  those of the
individual fullerenes.  Therefore, the natural first step is to specify the
model which we employ for each buckyball.  For simplicity and ease of
mathematical analysis, but at the same time retaining essential geometrical
characteristics, we assume that an electron gas is confined to the surface
of a sphere of chosen radius \cite{Inaoka,Devreese,Yannouleas}.
This spherical two-dimensional electron  gas
(S2DEG) is characterized by the electron effective mass $m^\ast$ and  the
number $N_F$ of occupied energy levels.  Our model enables us to exploit
spherical symmetry of the particulates  in the Bloch-Floquet  theorem for
generating the  wave functions in a linear array \cite{aizin}.    This effective mass model
is suitable for ``small" nano spheres, i.e., when the separation between energy
levels is large.  Additionally, the energy band structure may be included into
our formalism through form factors for the Coulomb matrix elements and the
polarization function. In this regard, we note that
electron energy loss spectroscopy (EELS) has been used to probe
the plasmon excitations for concentric-shell fullerenes
embedded  in a film \cite{EELS}. Furthermore,  perfectly spherical
 shells were used in the theoretical modeling of the EELS data
and the agreement was good.  The model of Lucas, et al. \cite{Lucas}.
was shown to be qualitatively adequate for understanding the optical
data for multi-shell fullerenes. In that work \cite{Lucas},
 the ultraviolet dielectric tensor of monolayer graphene is adapted
 to  the
spherical geometry of a fullerene by averaging over the three
possible orientations of the ${\bf c}$ axis.  Thus, a continuum
model was used by Lucas, et al. \cite{Lucas} starting from the
planar local dielectric function of planar monolayer graphene.

\medskip

The rest of this paper is organized as follows. Section II contains our theoretical framework
and the results based on it. Illustrative comparisons with data sets are given there as
well. The last section, Sec. III, is devoted to a short summary and relevant comments.

\section{General Formulation of the Problem}
\label{sec2}

Let us consider  a linear array of spherical
two-dimensional electron gases,
 with their centers
located on the $x$ axis. The center of each shell is
at $x=na$ ($n=0,\pm1,\pm2,\cdots$). Each
``ball" consists of $N$ concentric shells  with radii
$R_1<R_2<\cdots<R_N$, where $a>2R_N$. For simplicity, we assume
that each spherical shell is infinitesimally thin.
 We will construct the electron wave functions in the form of Bloch
combinations as described by Huang and Gumbs for an array of
rings \cite{huangandgumbs} and by Gumbs and Aizin for an array
of cylinders \cite{aizin}. In the absence of tunneling
between the shells, the single-particle
Bloch wave functions for the  array
with the periodicity of the lattice are given by
\begin{equation}
<{\bf r}|\nu>=\frac{1}{N_x}
\sum_{j=-\frac{N_x}{2}}^{\frac{N_x}{2}}e^{ik_xja}
\Psi_{lm} (\vec{r}-ja\hat{e}_x)
\ ,\ \ \ \Psi_{lm}(\vec{r}-ja\hat{e}_x)=
\frac{\mathcal{R}_j(r)}{R}
Y_{lm}(\Omega),\ \ \
\label{gae33}
\end{equation}
where $\nu=\{k_x,l,m\}$ is a composite index
for the electron eigenstates, $\Psi_{lm}(\vec{r})$  is
the wave function for an  electron, with
 angular momentum quantum
numbers $l=0,1,2,\cdots$ and $|m|\leq l$,
$\mathcal{R}_j^2(r)=\delta(r-R_j)$. Additionally, $k_x=\frac{2\pi }{L_x}n$ with
$n=0,\pm1,\pm2,\cdots,\pm\frac{N_x}{2}$. Here,  $N_x=L_x/a$ is the number
 of nano balls
in the array with periodic boundary conditions. Electron motion in the
azimuthal direction around the shells is quantized and characterized by
the angular momentum quantum number $l$. The electron spectrum
in each shell is discrete and given by $\epsilon_{\nu}=\hbar^2 l(l+1)/2\mu^\ast R^2$ . The
spectrum does not depend on $m$ and $k_x$.
\par

Plasmons may be obtained from the solution of the density matrix
equation, as described above. We have

$\displaystyle{i\hbar\frac{\partial\hat{\rho}}
{\partial t}=[\hat{{\cal H}},\hat{\rho} ]}$ \cite{aizin}.
For $\hat{{\cal H}}=\hat{{\cal H}}_0-e\tilde{\varphi}$ and $\hat{\rho}
=\hat{\rho}_0+\delta\hat{\rho}$, with $<i\nu|\hat{{\cal H}}_0|i^\prime\nu
^{\prime}>
=\epsilon_{i\nu}\delta_{\nu\nu^{\prime}}\delta_{ii^\prime}$,
$<i\nu|\hat{\rho}_0|i^\prime\nu^{\prime}>
=2f_0(\epsilon_{i\nu})\delta_{\nu\nu^{\prime}}\delta_{ii^\prime}$, we obtain
in the lowest order of perturbation theory
\begin{equation}
<i\nu|\delta\hat{\rho}|i^{\prime}\nu^{\prime}>=2e\
 \frac{f_0(\epsilon_{\nu})-f_0(\epsilon_{\nu^{\prime}})}
{\hbar\omega-\epsilon_{\nu}+\epsilon_{\nu^{\prime}}}
<i\nu|\tilde{\varphi}({\bf r})|i^{\prime}\nu^{\prime}>
\ ,
\label{gae34}
\end{equation}
where $f_0(\epsilon)$ is the Fermi function and $\tilde{\varphi}({\bf r})$
is the induced potential. The potential $\tilde{\varphi}({\bf r})$
satisfies Poisson's equation
\begin{equation}
\nabla^2\tilde{\varphi}({\bf r})=\frac{4\pi e}{\epsilon_s}
\delta n({\bf r},\omega)\ ,
\label{gae35}
\end{equation}
where  $\delta n({\bf r},\omega)$ is the fluctuation of electron density.
Making use of the relation
\begin{equation}
\delta n({\bf r},\omega)=\sum_{ii^\prime}\sum_{\nu,\nu^{\prime}}
<{\bf r}|i\nu><i\nu|\delta\hat{\rho}|i^{\prime}\nu^{\prime}><i^{\prime}\nu^{\prime}|{\bf r}>
\label{gae36}
\end{equation}
and Eq.\ (\ref{gae34}), we may write in Fourier representation
\begin{equation}
\delta n({\bf q},\omega)=\frac{2e}{V}\ \sum_{\nu,\nu^{\prime}}
 \frac{f_0(\epsilon_{\nu})-f_0(\epsilon_{\nu^{\prime}})}
{\hbar\omega-\epsilon_{\nu}+\epsilon_{\nu^{\prime}}}
<i^{\prime}\nu^{\prime}|e^{-i{\bf q}\cdot{\bf r}}|i\nu>
\sum_{{\bf q}^{\prime}}\tilde{\varphi}({\bf q}^{\prime})
<i\nu|e^{i{\bf q}^{\prime}\cdot{\bf r}}|i^{\prime}\nu^{\prime}>
\ ,
\label{gae37}
\end{equation}
where $\delta n({\bf q},\omega)$ and $\tilde{\varphi}({\bf q})$ are 3D
Fourier transforms of $\delta n({\bf r},\omega)$ and $\tilde{\varphi}({\bf r})$,
respectively and ${\bf q}=(q_x,q_y,q_z)$. The matrix elements
$<i\nu|e^{i{\bf q}\cdot{\bf r}}|i^\prime\nu^{\prime}>$
with wave functions $|i\nu>$ given in Eq.\ (\ref{gae33})
may be evaluated as follows
\begin{equation}
<i\nu|e^{i{\bf q}\cdot{\bf r}}|i^{\prime}\nu^{\prime}>= 4\pi\
 \delta_{ii^{\prime}}\delta_{k_x^{\prime},k_x-q_x+G_N}
\ \sum_{L,M} \ \ i^L j_L(qR_i)   Y_{LM}^\ast(\hat{\bf{q}})
\int d\Omega\ \  Y^{\ast}_{lm}(\Omega)
Y_{LM}(\Omega)\
Y_{l^{\prime}m^{\prime}}(\Omega)
\ ,
\label{gae38}
\end{equation}
where $G_N=2\pi N/a$ with $N=0,\pm1,\pm2,\cdots$.
Substituting Eq.\ (\ref{gae38}) into Eq.\  (\ref{gae37}),
we obtain after some straightforward algebra

\begin{eqnarray}
&&\delta n({\bf q})=\frac{8\pi e}{V }\frac{L_x}{a}\sum_{i}\
\sum_{L}
\sum_{l,l^\prime}
\frac{f_0(\epsilon_{l})-f_0(\epsilon_{l^\prime})}
{\hbar\omega+\epsilon_{l^\prime}-\epsilon_{l}}
\ (2l+1)(2l^\prime+1)
\left( \begin{matrix}
l&l^\prime& L\cr
 0 & 0 & 0\cr
\end{matrix}\right)^2
\nonumber\\
&&\times    \sum_{M}\  \   j_L(qR_i)
Y_{LM}(\hat{\bf{q}})\
\nonumber\\
&&\times
\sum_{N=-\infty}^{\infty}\sum_{q_y^\prime ,q_z^\prime}
\ \tilde{\varphi}
\left(q_x+G_N,q_y^\prime,q_z^\prime\right)
 Y_{L M} (\hat{\bf{q}}_N^\prime)
 j_L\left(\sqrt{(q_x+G_N)^2 +q_y^{\prime\ 2}
 +q_z^{\prime\ 2}}\right)
 \ .
\label{gae39}
\end{eqnarray}

The potential $\tilde{\varphi}(q)$ may be written in terms of
$\delta n({\bf q},\omega)$
as $\tilde{\varphi}(q)=-4\pi e \delta n({\bf q},\omega)/\epsilon_sq^2$.
Using this relation in Eq.\  (\ref{gae39}), we obtain
\begin{equation}
\delta n({\bf q},\omega)=-\frac{32\pi^{2} e^2}{a^{2}\epsilon_s}
\sum_{L,M}\Pi_{L}(\omega)
Y_{LM}(\hat{\bf{q}})j_L(qR)\
U_{L,M}(q_x)\ ,
\label{gae40}
\end{equation}
where $\Pi_{L}( \omega)$ is the susceptibility function in a
single spherical shell of radius $R$ given by
\begin{equation}
\Pi_{L}(\omega)=  \sum_{l,l^\prime}
\frac{f_0(\epsilon_{l})-f_0(\epsilon_{l^\prime})}
{\hbar\omega+\epsilon_{l^\prime}-\epsilon_{l}}
(2l+1)(2l^\prime+1)
\left( \begin{matrix}
l&l^\prime& L\cr
 0 & 0 & 0\cr
\end{matrix}\right)
^2
\label{gae13}
\end{equation}
and
\begin{eqnarray}
U_{L,M}(q_x)&=&\frac{1}{L_yL_z}\sum_{N=-\infty}^{\infty}
\sum_{q_y,q_z}
\frac{\delta n(q_x+G_N,q_y,q_z,\omega)}{\left(q_x+G_N\right)^2+q_y^2+q_z^2}
j_L\left(\sqrt{\left(q_x+G_N\right)^2+q_y^2+q_z^2}R \right)
\nonumber\\
&\times &
Y_{LM}^{\ast}\left(\frac{(q_x+G_N,q_y,q_z)}
{\sqrt{(q_x+G_N)^2+q_y^2+q_z^2}} \right)
\ .
\label{gae41}
\end{eqnarray}

Substituting the expression for $\delta  n({\bf q})$ given in
Eq.\ (\ref{gae40})
into Eq.\ (\ref{gae41}), we obtain a set of linear equations which have nontrivial
solutions provided their determinant is zero,


\begin{equation}
\sum_{L,M}\left[\delta_{L L^{\prime}}\delta_{M M^{\prime}}+\frac{8e^2}{a\varepsilon_s}\Pi_{L}\left(\omega\right)V_{L^\prime M^\prime,LM}\left(q_x, a \right)\right]U_{LM}\left(q_x,\omega\right)=0
\label{gae42}
\end{equation}
with $L,L^{\prime}=1, 2, 3,\cdots$ and $M,M^{\prime}
     =0,\pm1,\pm2,\cdots,\pm L$. Also, we have
\begin{equation}
\label{eq2}
V_{L^\prime M^\prime, LM}\left(q_x, a \right)=\sum\limits_{N=-\infty}^{\infty}
V_{L^\prime M^\prime, LM}^{\left(N\right)}\left(q_x, a \right),\\
\end{equation}
is the matrix for the Fourier transform of the Coulomb interaction potential
between electrons on the spherical shells and
\begin{eqnarray}
V_{L^\prime M^\prime, LM}^{\left(N\right)}\left(q_x,a \right)&
=&\int_{-\infty}^{\infty} dq_y\int_{-\infty}^{\infty} dq_z
\frac{j_L\left(q_N R\right)j_{L^\prime}\left(q_N R\right)}{q_N^2}
Y_{L^\prime M^\prime}^{\ast}\left(\hat{\bf q}_N\right)Y_{LM}\left(\hat
{\bf q}_N\right),\label{eq3}\\
{\bf q}_N&=&\left(q_x+N\frac{2\pi}{a}\right){\bf i}+q_y{\bf j}+q_z{\bf k} \ .
\end{eqnarray}
It follows that  $V_{L^\prime M^\prime, LM}\left(q_x,a \right)
=V_{LM,L^\prime M^\prime}^{\ast}(q_x,a )$ and therefore we have only six
independent matrix elements for the Coulomb interaction. when $L=L^\prime=1$
and $M,M^\prime=0,\pm 1$. Equation~(\ref{eq2})    gives the Coulomb interaction
matrix elements in between two electron states with quantum numbers $\{L,M\}$
and $\{L^\prime,M^\prime\}$.
In the case when the external probe uses circularly polarized light,
only  $L=L^\prime=1$ and $M,M^\prime=0,\pm1$ are included in the dispersion
equation.


Equation (\ref{gae42}) determines the dispersion equation for
the collective plasmon  excitations. The frequencies of these
excitations are dispersive, unlike the case for a single shell,
a pair of concentric shells, or a pair of non-overlapping shells
not sharing a common center. Additionally, a new feature is that
for the superlattice, the plasmon modes not only depend on $L$ but $M$
as well.  At $T=0$ K, it is a straightforward matter to evaluate
numerically the susceptibility function $\Pi_{L}(\omega)$.
Equation (\ref{gae42}) shows that the symmetry of
the lattice is maintained in the dispersion equation and that
the plasmon excitations depend on the wave vector
$q_x$ in the $x$ direction with period $G=2\pi/a$ as well as the
wave vector $q_y$. At this point, it should be clear that there is no
change in the formal expression  for the dispersion equation arising from
(\ref{gae42}) by interchanging $q_x$ and $q_z$. However,  the numerical values
for the Coulomb matrix elements are not equal axis of quantization  is in the
same direction or perpendicular to the axis of quantization.
\medskip

In the limit $a\to\infty$, the sum over  reciprocal lattice vectors in
Eq.\ (\ref{gae42}) gets transformed into an integral, and the
determinantal matrix in Eq.\ (\ref{gae42}) becomes diagonal in
the indices $L$ and $L^\prime$ by using the result

\begin{equation}
\overline{V}_{LM,L^\prime M^\prime}\left(R \right)\equiv
\int  \frac{d^3\textbf{q}}{q^2} \ Y_{LM}(\hat{\textbf{q}})j_L(qR)
Y_{L^\prime M^\prime}(\hat{\textbf{q}})j_{L^\prime}(qR)
=\frac{\pi}{2(2L+1)R}\ \delta_{LL^\prime}\delta_{MM^\prime} .
\label{gae43}
\end{equation}
In this limit, we obtain the following  equation to solve for plasmons, i.e.,

\begin{equation}
\prod_{L=1,2,\cdots}\mbox{Det}\left[1+
\frac{2e^2}{ \epsilon_s}\frac{1}{(2L+1)R}\Pi_{L}(\omega)\right] \delta_{LL^{\prime}} =0,\
\label{gae44}
\end{equation}
by making use of the orthogonality of spherical harmonics and the relation
$\int_0^\infty dx\ j_L^2(x)=\pi/2(2L+1)$ for  spherical Bessel functions.
Equation (\ref{gae44}) shows that the angular momentum quantum numbers are
completely decoupled and the  plasmon equation for a single S2DEG depends
on the angular momentum quantum number $L$ and not on its projection $M$ on the axis
of quantization \cite{Inaoka,Devreese,Yannouleas}.  Furthermore, these angular
momenta are decoupled from one  another so that $L$ is a good quantum number
for labeling plasmons on isolated shells.
\medskip

In the next section, we solve Eq.\ (\ref{gae42}) numerically for
$L,L^\prime= 1$ and $M,M^\prime=0,\pm 1$. This would correspond to
an external perturbation using circularly polarized light.
The higher angular momentum with $L>1$ may be achieved by a special
light beam, such as, a helical light beam.
However, when two S2DEGs have their centers well separated so
that the Coulomb coupling is negligible, a non-circularly polarized
light source may excite several modes.

\medskip

\section{Numerical Results}
\label{sec4}

We now present our numerical results for the cases that the microribbon   lies
 either parallel to the $z$ axis or the $x$ axis. All calculations were carried
 out at zero temperature. The radius of each spherical shell
  was taken to be $R=10$ nm and the angular momentum quantum number for the highest
  occupied state at the Fermi level for each S2DEG was chosen as     $l_F=10$.
The corresponding Fermi energy is 0.168 eV.

\begin{figure}[t] 
  \centering
  \includegraphics[width=5.67in,height=3.84in,keepaspectratio]{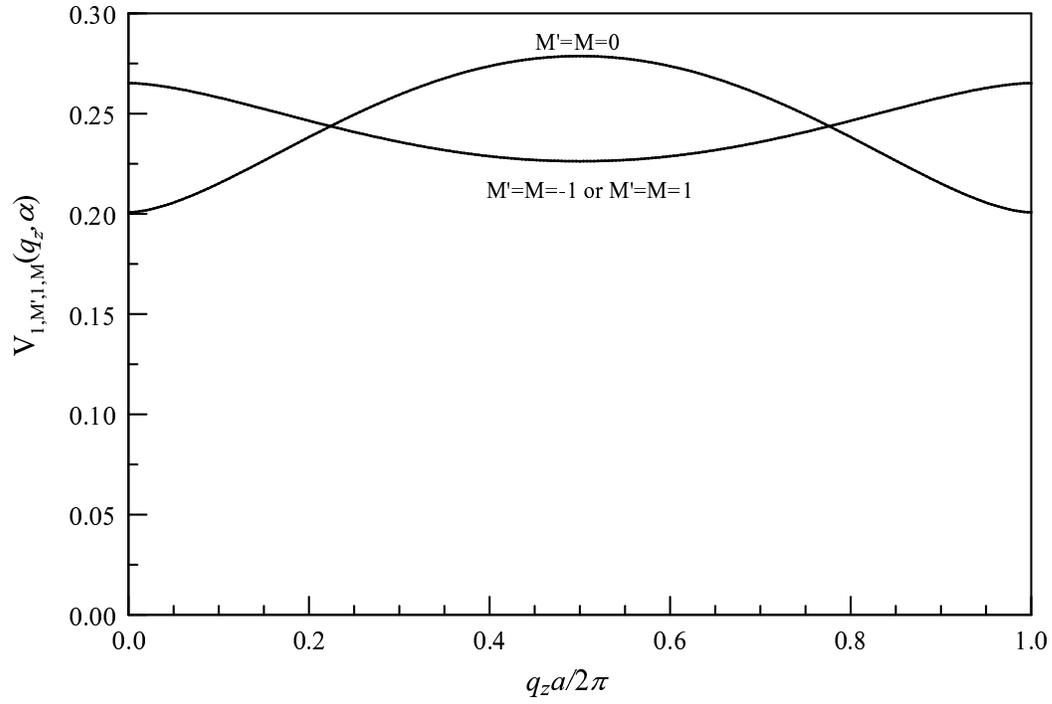}
  \caption{Non-zero Coulomb interaction matrix elements for a linear chain
  of spherical shells along the $z$ axis when the separation between the
   centers of consecutive spheres is chosen as $a=3R$, and $L=L^\prime=1$
   with  $M,M^\prime=0,\pm 1$.}
  \label{fig1}
\end{figure}

In Fig.~(\ref{fig1}), we plotted the non-zero elements of the Coulomb interaction matrix
as functions of the dimensionless wave vector $q_xa /2\pi$ for a linear
 chain of shells that lie along the $z$ direction. Only the diagonal matrix elements
corresponding to  $M^\prime=M=0,\pm 1$ are finite. The off diagonal matrix elements are
zero so there is no Coulomb interaction between electrons of different azimuthal
quantum number. We note that there are crossings  for the interaction matrix
elements    at the points  $q_x\alpha/2\pi=0.23$ and $0.77$.

\begin{figure}[ht] 
  \centering
  \includegraphics[width=5.67in,height=3.85in,keepaspectratio]{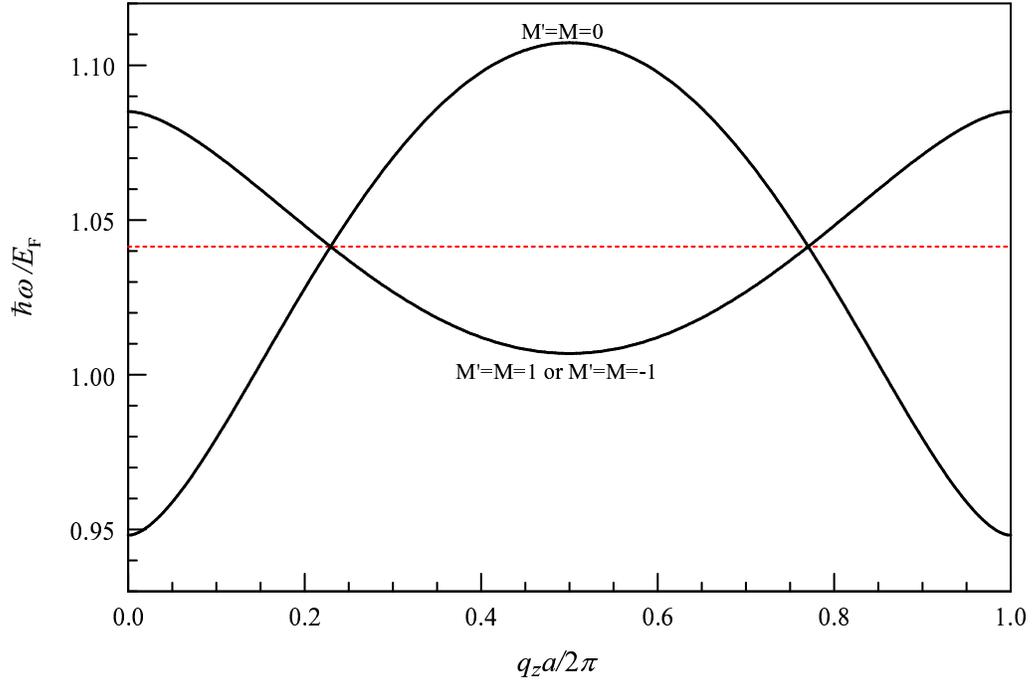}
  \caption{Plasmon dispersion relation in the first Brillouin zone for a periodic array
   of S2DEGs along the $z$ axis with period    $a=3R$.
The radius of each sphere is  $R=10$ nm and $l_F=10$ is the angular momentum
quantum number for the highest occupied electron state on each sphere.
The horizontal  solid line corresponds to  the plasmon energy for the single shell   with $L=1$. }
  \label{fig2}
\end{figure}

We next plot in Fig.~(\ref{fig2}), our results for  the plasmon dispersion
 relation    in the first Brillouin zone for a periodic array  of S2DEGs
 extended along the $z$ axis. We see that there are two plasmon  branches,
 one  corresponding to $M^\prime=M=0$ and a  degenerate one which corresponds
 to the case where $M^\prime=M=\pm 1$. The two branches cross at the same points
 where the crossing occurs in Fig.~(\ref{fig1}). There is no interaction between
 them due to the zero value of the off diagonal matrix elements. The horizontal
 solid line in the figure shows the energy of the single shell plasmon for $L=1$.

\begin{figure}[t] 
  \centering
  \includegraphics[width=5.67in,height=3.82in,keepaspectratio]{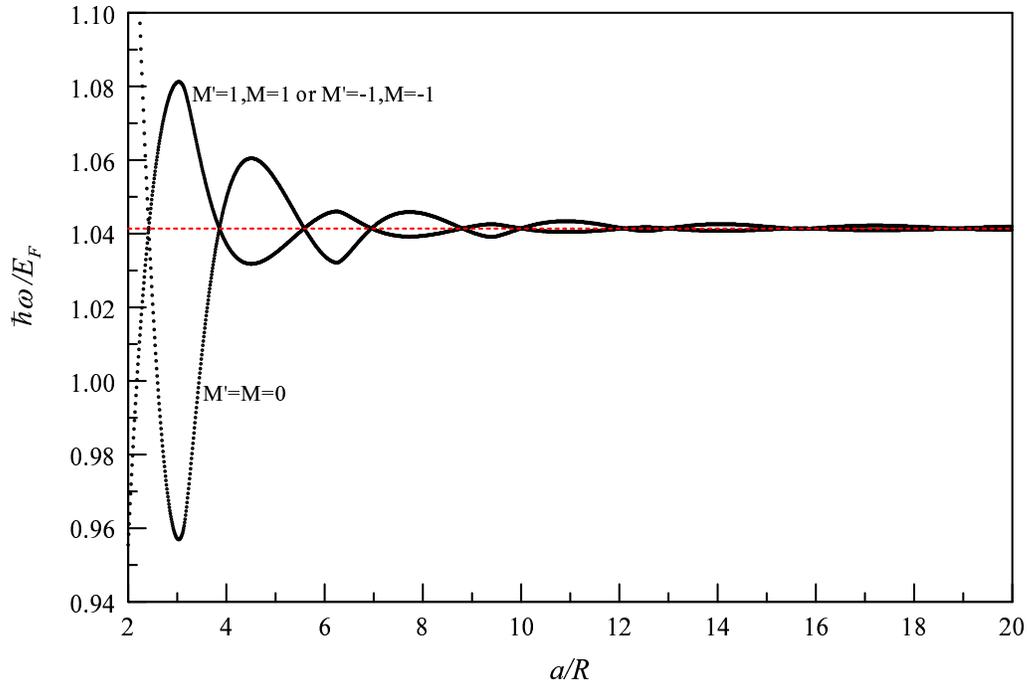}
  \caption{Plasmon variation for fixed $q_z R=2$ as a function of the
  separation $a$ of the shells along the $z$ axis.The horizontal
    dotted line gives the single shell plasmon energy for $L=1$.}
  \label{fig3}
\end{figure}

  Figure~(\ref{fig3}) displays the variation  of  the plasmon  frequencies
  with the separation $a$    along the $z$ direction.  The  wave vector is
  fixed given by  $q_z R=2$. We note that we have oscillations of the plasmon
  energies with decaying amplitude as the separation increases. The horizontal
  dotted line indicates the single shell plasmon energy for $L=1$. The oscillatory
  behavior of the plasmon branches shows that for some range of separation
  between shells the repulsive Coulomb  interaction dominates over the negative
  exchange energy, thereby resulting in a larger plasmon energy than that for
  a single S2DEG. In other ranges of  separation $a$, it is the exhange interaction
  which makes the   dominant contribution, resulting in the plasmon oscillations.

\begin{figure}[t] 
  \centering
  \includegraphics[width=5.67in,height=3.87in,keepaspectratio]{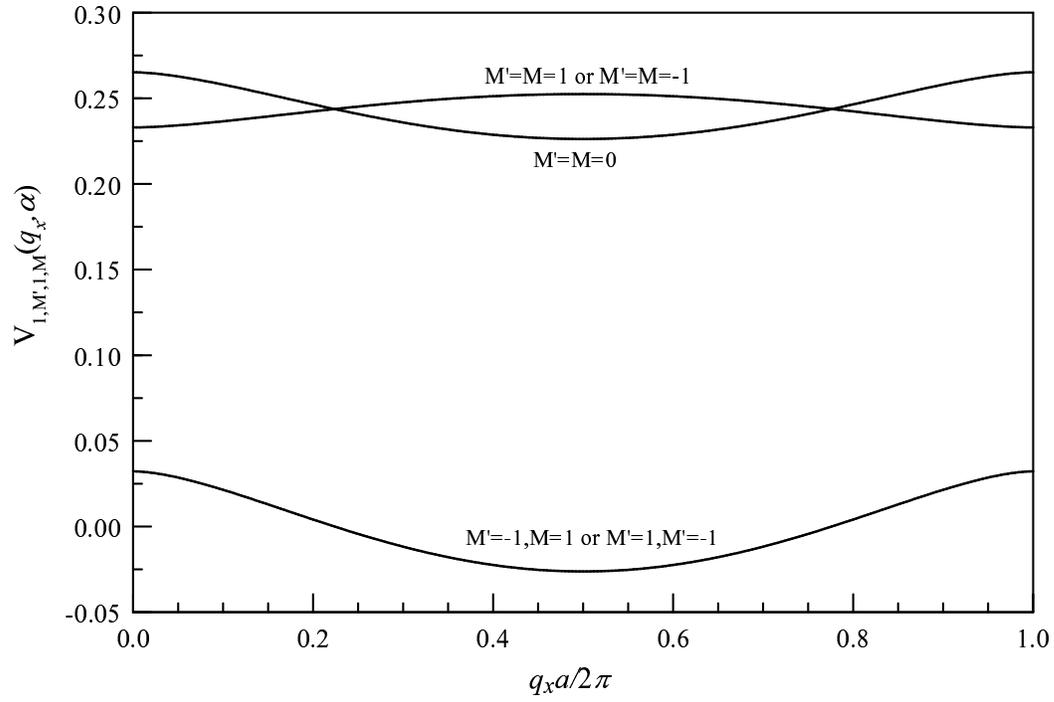}
  \caption{Coulomb interaction matrix elements for a linear array of shells along
  the $x$ axis for the case when $L=L^\prime=1$ and $M,M^\prime=0,\pm 1$.
  The radius of each sphere is  $R=10$ nm. }
  \label{fig4}
\end{figure}

 In Fig.~(\ref{fig4}), we plot the five non-zero Coulomb interaction matrix elements
 as functions of  the dimensionless wave vector $q_x a /2\pi$ for a periodic array
  of S2DEGs  whose centers lie on  the $x$ axis. Contrary to the case  of the fullerenes
  along the $z$  axis there are now nonzero off diagonal matrix elements
  $M^\prime\neq M$. Comparing Figs.~(\ref{fig3}) and (\ref{fig4}) we conclude that
  there is anisotropy in the Coulomb interaction energy between electrons based in the
  orientation of the fullereness.

\begin{figure}[t] 
  \centering
  \includegraphics[height=3.87in,keepaspectratio]{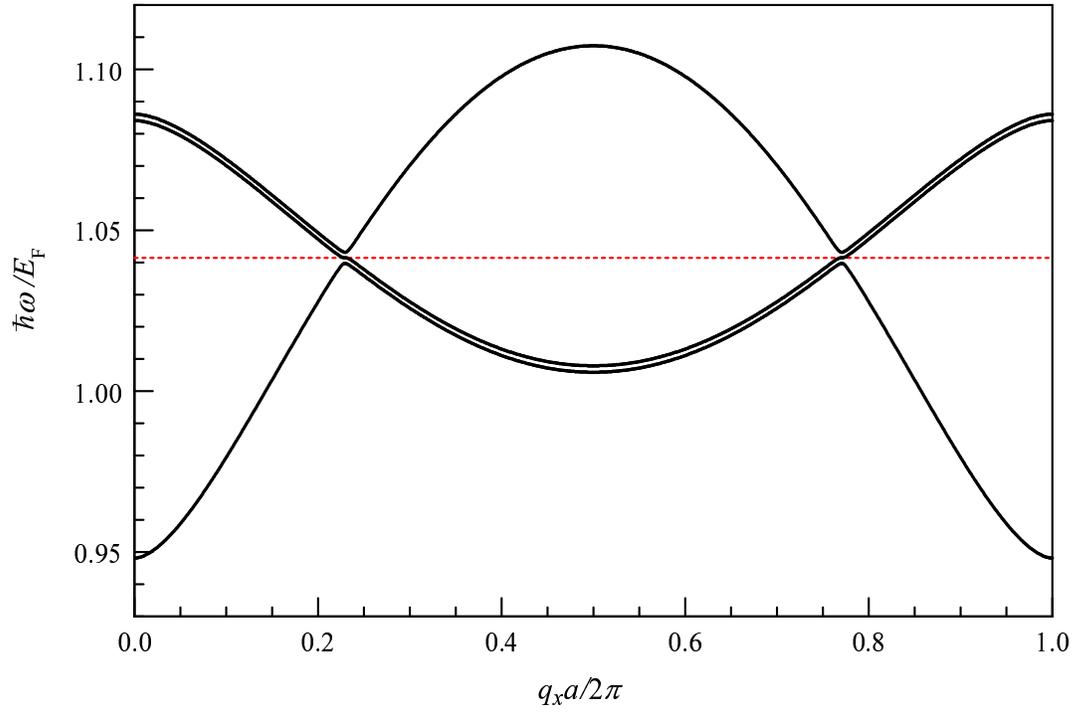}
  \caption{Plasmon dispersion relation  for a linear array of spherical
  shells of radius $R=10$ nm separated by a distance $a=3R$ along the $x$ axis.}
  \label{fig5}
\end{figure}

\begin{figure}[t] 
  \centering
  \includegraphics[height=3.87in,keepaspectratio]{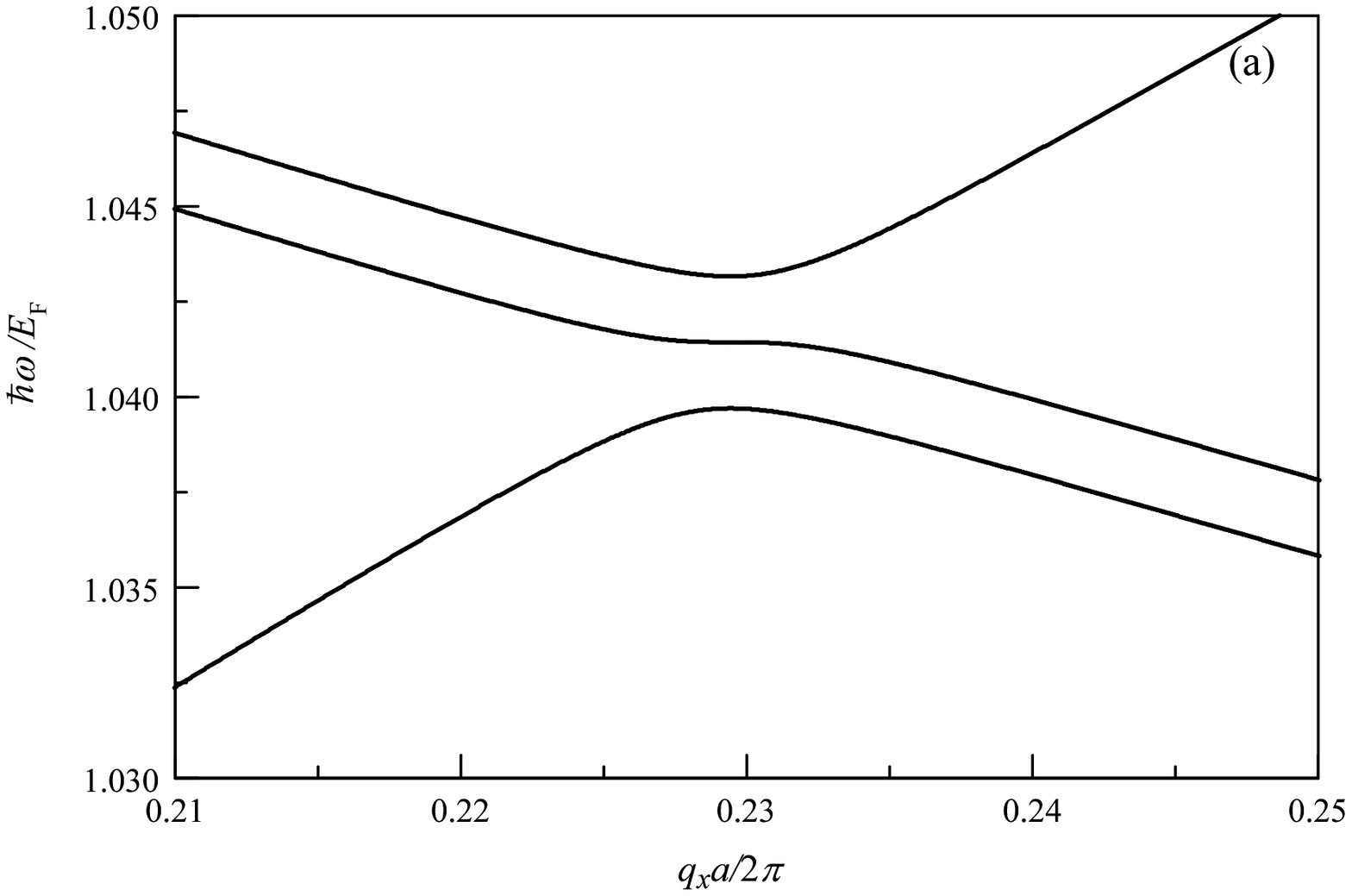}
  \includegraphics[height=3.87in,keepaspectratio]{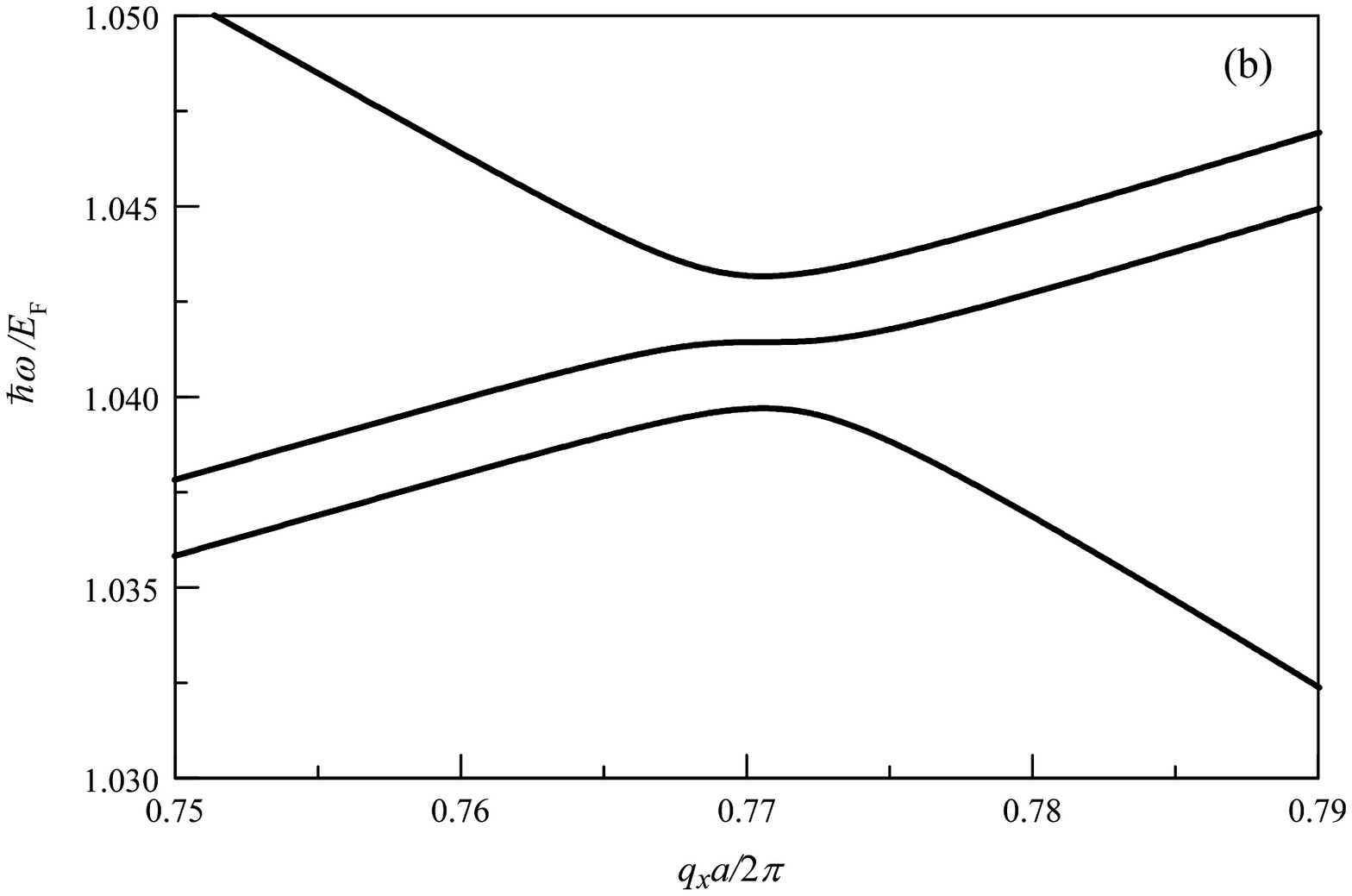}
  \caption{Anticrossing of the plasmon modes of Fig.~\ref{fig5} in the
  vicinity of the points (a)  $q_x/2\pi a=0.23$ and (b)      $q_x/2\pi a=0.77$.}
  \label{fig6}
\end{figure}

\begin{figure}[t] 
  \centering
  \includegraphics[width=5.67in,height=3.85in,keepaspectratio]{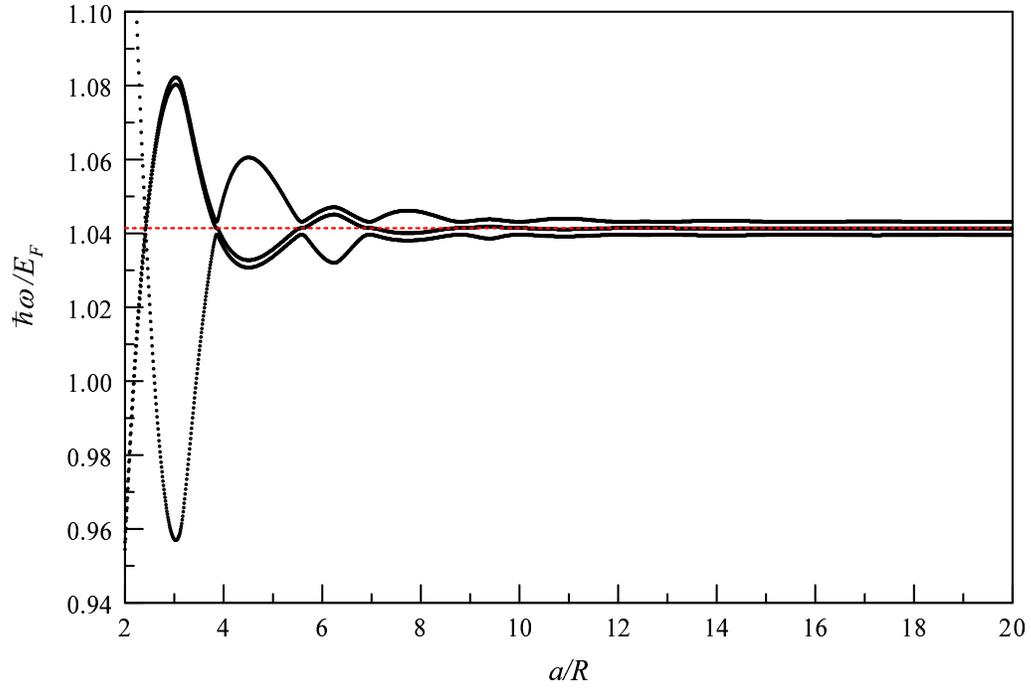}
  \caption{Variation of the  plasmon excitation energies with separation between adjacent
   S2DEGs on a linear chain on the $x$ axis  for chosen $q_x R=2$.  The horizontal
   dashed line corresponds to the plasmon mode frequency for a S2DEG in isolation. }
  \label{fig7}
\end{figure}

We present in  Fig.~(\ref{fig5}) the results of our calculations for the
plasmon excitation energy for a superlattice  of   S2DEGs extending along the $x$
axis. There are now three plasmon branches instead of two which we obtained in Fig~(\ref{fig2}),
due to the lifting of the degeneracy by the off diagonal Coulomb matrix elements. The extra
plasmon mode originates  from the interaction between electrons with $M^\prime=-1,M=1$
or $M^\prime=1,M=-1$. We see in the same figure that there is a strong interaction between
plasmon modes at the points $q_x\alpha/2\pi=0.225$ and $0.775$ which leads to the opening of
a gap in the plasmons energy. This effect of anticrossing of the branches is more
clearly presented in Figs.~(\ref{fig6}a) and (\ref{fig6}a).
The dependence of the plasmon  energies on the separation between the centers
of adjacent S2DEGs on a linear chain along the $x$-axis is shown in   Fig.~(\ref{fig7}).
The oscillations in the plasmon branches exhibited in  Fig.~(\ref{fig7}) arise
as a consequence of the competition  between the direct Coulomb interaction and the
exchange interaction between spheres, as we explained for the results in Fig.~(\ref{fig3}).
The anticrossings appear at several values of the inter-sphere separation and are
due to the presence of the Coulomb interaction between shells.
Furthermore, these results show the   significance of the Coulomb interaction
for the superlattice compared with  the   plasma mode frequencies for the
single shell   for the chosen range. We notice     that the anticrossings
in  Fig.~(\ref{fig7})   occur at the single shell plasmon energy. This is where the
repulsive Coulomb and attractive exchange interactions between shells
for two plasmon branches cancel each other.
We chose  $q_x R=2$ in these calculations but  the described   characteristics are not
restricted to our choice of wave  vector.

\section{Concluding Remarks}
\label{sec4}

This paper is  devoted to a model calculation of the plasmon dispersion relation
for a narrow microribbon of fullerene atoms or metallic shells which we simulate by a linear  periodic
array of S2DEGs.  The neglect of the width of the ribbon means that our study
does not investigate edge effects.  The work   explores the coupling between
plasmon excitations in the 2D electron gases  occupying the surfaces of
 an infinite number of  spheres.
By adopting the quantization axis to be perpendicular to the inter-particle
 axis of the chain,   we were able to calculate the
 Coulomb matrix elements, resulting  in a strong anisotropy in the plasmon coupling
 with respect to the direction of the probe field. While the applications
to fullerenes have been  emphasized, we believe that the results also have broader
 implications to   metallic particulates. The data we obtained  reveal significant new information
for the area of plasmonics.   Our result on the spatial correlation may
 be experimentally observable. In this connection,  there have already been
some experimental reports pointing to a similar effect in nanoparticles \cite{Yang1}.
Additionally, it seems that there exists a relation between this work and the theory of
 hybridization for surface plasmons in metallic dimers \cite{Nordlander,GG2014}
 with regard to whether the    quantization axis
 is parallel or perpendicular to   the inter-particle axis.

There have been several examples when anisotropic properties of condensed matter 
systems have  been demonstrated. Among these, we mention the  electrical, thermal, mechanical
and chemical properties of graphite along the $a,\ b$ and $c$ directions
\cite{n21} as well as elastic properties of carbon nanotube bundles \cite{n22}. In addition, the dispersion
relation of the high energy optical $\pi$-plasmons for graphite was calculated by Chiu, et al., \cite{n23} who
showed that the plasmon frequency depends on whether the momentum transfer is parallel or perpendicular
to the hexagonal plane within the Brillouin zone. The anisotropic conductivity of epitaxial graphene on
SiC was presented in Ref.[\onlinecite{n24}].
There have been attempts to exploit the anisotropy of these properties to device applications.
For example, the  authors  of Ref.\,[\onlinecite{n42}] explored the possibility of employing 
the tuning of surface plasmon frequencies to more efficient optical sensors.

\begin{acknowledgments}
This research was supported by  contract \# FA 9453-13-1-0291 of  of
AFRL.
\end{acknowledgments}


\newpage

\end{document}